\documentclass[aps, prd,twocolumn,floatfix,amsmath,superscriptaddress,nofootinbib]{revtex4}

\usepackage{bm}
\usepackage{epsfig}
\usepackage{latexsym}
\usepackage{amsmath}
\usepackage{amssymb}
\usepackage{amsfonts}
\usepackage{ifpdf}

\usepackage[mathscr,scaled=1.15]{urwchancal}
\DeclareFontFamily{OT1}{pzc}{}
\DeclareFontShape{OT1}{pzc}{m}{it}%
{<-> s * [1.15] pzcmi7t}{}
\DeclareMathAlphabet{\mathpzc}{OT1}{pzc}{m}{it}

\usepackage{color}

\usepackage{supertabular}
\usepackage{placeins}
\usepackage{graphicx}

\usepackage{relsize}
\RequirePackage{xspace}

\definecolor{purple}{rgb}{0.5,0,0.5}
\definecolor{blue}{rgb}{0.0,0,0.9}
\definecolor{prdblue}{rgb}{0.133,0.118,0.498}
\def\bfsigma{\mbox{\boldmath $\sigma$}}

\usepackage[colorlinks=true, pdfstartview=FitV, linkcolor=prdblue, citecolor= prdblue, urlcolor=prdblue]{hyperref}
\usepackage{slashed}

\begin{document}

\title{Radiative decays of $\Upsilon(nS)$ into S-wave and P-wave charmonium}

\author{Dan-Dan Shen, \footnote{181002016@stu.njnu.edu.cn}
Chong-Yang  Lu, \footnote{181002013@stu.njnu.edu.cn}
Peng Sun,\footnote{06260@njnu.edu.cn }
Ruilin Zhu ~\footnote{Corresponding author:rlzhu@njnu.edu.cn} }
\affiliation{Department of Physics and Institute of Theoretical Physics,
Nanjing Normal University, Nanjing, Jiangsu 210023, China}

\date{\today}

\begin{abstract}
Motivated by very recent measurement of the radiative decays of $\Upsilon(1S)$ to $\chi_{c1}$ at Belle, we use the nonrelativistic QCD factorization theory and calculate the branching fractions of the radiative decays of bottomonium into S-wave and P-wave charmonium, i.e. $\Upsilon(nS) \to \eta_c(nS)+\gamma$ and $\Upsilon(nS)\to \chi_{cJ}+\gamma$.  We systematically studied
the branching fractions  of the radiative decays of bottomonium into charmonium. Compared to the previous calculation, we obtained the analytical expression for the decay widths and considered the color-octet contributions. For $\Upsilon(nS) \to \eta_c(nS)+\gamma$, the relativistic corrections are also obtained.  Through the calculation, the theoretical prediction for  $\Upsilon(1S)\to\chi_{c1}+\gamma$ is still smaller than the recent Belle measurement. Further theoretical work and experimental analysis are necessary to understand the $\chi_{c1}$ production mechanism in upsilon decays.
\end{abstract}


\maketitle


\section{Introduction}\label{introduction}
%
Heavy quarkonium production and decay are good channels to test the hadronization mechanism and also understand the nonrelativistic Quantum Chromodynamics (NRQCD) factorization theory. Very recently, the Belle Collaboration has reported the first observation of the radiative decays of $\Upsilon(1S)$ to $\chi_{c1}$ based on a data sample collected at the $\Upsilon(2S)$ energy with an integrated luminosity of $24.9fb^{-1}$~\cite{Katrenko:2019vdd}. For the channels of $\Upsilon(1S)$  to other charmonium, the upper limits for their branching fractions are also given~\cite{Katrenko:2019vdd,Shen:2010iu}, see Tab.~\ref{belle}.
\begin{table}
\caption{Experimental measurements of branching fractions (in units of $10^{-5}$) of the radiative decays of bottomonium into charmonium.
}\label{belle}
\begin{tabular}{c|cc}
 \hline\hline
Decay channel  & Belle-2019~\cite{Katrenko:2019vdd}  & Belle-2010~\cite{Shen:2010iu}   \\
 \hline
 $\Upsilon(1S)\to \chi_{c1}+\gamma$ & $4.7_{-1.8-0.5}^{+2.4+0.4}$ & $<2.3$  \\

 $\Upsilon(1S)\to \chi_{c0}+\gamma$ & $<6.6$ & $<65$   \\

  $\Upsilon(1S)\to \chi_{c2}+\gamma$ &$<3.3$ ~& $<0.76$    \\
     $\Upsilon(1S)\to \eta_{c}(1S)+\gamma$ & $<2.9$ ~& $<5.7$  \\
$\Upsilon(1S)\to \eta_{c}(2S)+\gamma$  & $<40$ ~& $-$\\
    \hline\hline
\end{tabular}
\end{table}

As two classes of well-defined and well-established heavy quarknia, it is interesting to study the decay properties of bottomonium to charmonium in both experimental and theoretical aspects~\cite{Brambilla:2010cs}. The inclusive decays of $\Upsilon(1S)$ to $J/\psi$ were first observed by CLEO in 1989~\cite{Fulton:1988ug} and later measured by 35 times previous data sample corresponding to $21.6\times 10^6$  $\Upsilon(1S)$ decays in 2004~\cite{Briere:2004ug}. In 2016, the Belle collaboration measured ${\cal B}(\Upsilon(1S)\to J/\psi+\mathrm{anything})=(5.25\pm0.13(stat.)\pm0.25(syst.))\times 10^{-4}$ utilizing a data sample corresponding to $102\times 10^6$  $\Upsilon(1S)$ decays. The triple-gluon and single-photon decay modes are studied theoretically, which are not enough to explain the data~\cite{He:2009by}. Recent study indicates the single-gluon mode to produce $J/\psi$, i.e. the color-octet mechanism is important in the inclusive decays of $\Upsilon(1S)$ to  $J/\psi$ and may eliminate the difference between the theoretical prediction and experimental data~\cite{He:2019rwt}. The recent experimental review of
$\Upsilon(nS)$ physics can be found in Ref.~\cite{Jia:2020csg}.

Heavy quarkonium has multi-hierarchy of energy scales as $m_Qv^2 \ll m_Qv \ll m_Q$. By integrating out the scale of order $m_Q$ and higher momentum degrees of freedom, an effective field theory, i.e. NRQCD, has been established by Bodwin, Braaten and Lepage~\cite{Bodwin:1994jh}.
Within this framework, the production cross-section or decay width of heavy quarkonium can be factorized into two parts, the short distance coefficients and the long distance matrix elements(LDMEs). The short distance part can be calculated in powers of the strong-coupling constant $\alpha_s$ and the heavy quark relative velocity $v$ by perturbative theory, while the long distance part includes nonperturbative QCD dynamics and are process-independent. One big task is to test the universality of the NRQCD LDMEs when we obtain the short distance coefficients.

The exclusive decays of bottomonium into double charmonium  have been studied in Refs.~\cite{Jia:2007hy,Gong:2008ue,Braguta:2009xu,Braguta:2010zz}.  The inclusive decays of bottomonium into open charm or double charm baryon can be found in Refs.~\cite{Kang:2007uv,Zhang:2008pr,Sang:2012yh,Chen:2012zzg,Li:2020ggh}. The decays of $\Upsilon(nS)$ to exotic states can be found in Refs.~\cite{Zhu:2015jha,Zhu:2015qoa}. The radiative decays of bottomonium into charmonium are first studied in Refs.~\cite{Hao:2006nf,Gao:2007fv}. In this paper, we employ NRQCD factorization theory and calculate the branching fractions of the radiative decays of bottomonium into S-wave and P-wave charmonium, i.e. $\Upsilon(nS) \to \eta_c(nS)+\gamma$ and $\Upsilon(nS)\to \chi_{cJ}+\gamma$ with $J=0,1,2$. Both the $\gamma gg$ and single-photon decay modes will be considered. Compared to the calculation in
 Ref.~\cite{Gao:2007fv}, we have some improvements.  We will calculate the relativistic corrections in the process of $\Upsilon(nS) \to \eta_c(nS)+\gamma$. The analytical expression for the branching ratios will be given.
The color-octet mechanism will also be included when we consider the inclusive decays of upsilon into open charm and then  the charmonium is produced by the fragmentation of open charm.

This paper is organized as follows. In Sec.~II, we will show the main calculation formalism. Then, in Sec.~III and Sec.~IV, we will investigate the
color-singlet and color-octet contributions, respectively. In Sec.~V,  we will give the corresponding branching fraction  and do the phenomenological analysis. Finally, we will summarize.

\section{Formalism}
\label{SecII}
In this section, we will give the calculation formalism for the radiative decays of $\Upsilon(nS)$ into S-wave and P-wave charmonium.
\subsection{Parametrization of the decay amplitude}
The nontrivial amplitudes of $\Upsilon(nS)$ radiative decays into a charmonium can be written as
\begin{align}\label{LI}
&{\cal M}\left[\Upsilon(P,\epsilon)\to \eta_c(P_J)+\gamma(P',\epsilon^*)\right]\nonumber\\
=&ia\varepsilon^{\alpha\beta\mu\nu}\epsilon_\alpha(P) \epsilon^*_\beta(P') P_\mu P'_\nu,\\
&{\cal M}\left[\Upsilon(P,\epsilon)\to \chi_{c0}(P_J)+\gamma(P',\epsilon^*)\right]\nonumber\\=&b\epsilon(P) \cdot \epsilon^*(P')+c\epsilon(P) \cdot P' \,\epsilon^*(P')\cdot P,\\
&{\cal M}\left[\Upsilon(P,\epsilon)\to \chi_{c1}(P_J,\epsilon^*)+\gamma(P',\epsilon^*)\right]\nonumber\\=&id\varepsilon^{\alpha\beta\mu\nu}\epsilon^*_\alpha(P') \epsilon_\beta(P) \epsilon^*_\mu(P_J)  (P_\nu-P'_\nu)
\nonumber\\&+ie\varepsilon^{\alpha\beta\mu\nu}P_\beta \epsilon^*_\mu(P_J)(P_\nu-P'_\nu)  \epsilon^*_\alpha(P')\epsilon(P)\cdot P'
\nonumber\\&+if\varepsilon^{\alpha\beta\mu\nu}P_\beta \epsilon^*_\mu(P_J)(P_\nu-P'_\nu) \epsilon_\alpha(P)\epsilon^*(P')\cdot P,\\
&{\cal M}\left[\Upsilon(P,\epsilon)\to \chi_{c2}(P_J,\epsilon^*)+\gamma(P',\epsilon^*)\right]\nonumber\\=&
\epsilon^{*\alpha\beta}(P_J)[g\epsilon_\alpha(P) \epsilon^*_\beta(P')  +h\epsilon_\alpha(P)P_\beta\, \epsilon^*(P')\cdot P\nonumber\\
&+x\epsilon^*_\alpha(P')P_\beta\, \epsilon(P)\cdot P' +y P_\alpha P_\beta\, \epsilon^*(P)\cdot\epsilon^*(P')].\label{LI-4}
\end{align}
where we have the momentum conservation $P_J=P-P'$ and $\epsilon(P_H)$ is the polarization vector of the hadron $H$ with $\epsilon(P_H)\cdot P_H=0$. For the polarization tensor of $\chi_{c2}$, we also have $\epsilon^{\alpha\beta}(P_J)\cdot P_{J\alpha}=\epsilon^{\alpha\beta}(P_J)\cdot P_{J\beta}=0$. The  Lorentz invariants $a, b, c, d, e, f, g, h, x$, and $y$ rely on certain model or theory calculations.

\subsection{LDMEs}
In NRQCD, one can further write the decay width as the four-fermion LDMEs  with the corresponding perturbatively calculable short-distance coefficients $C_i(\mu)$
\begin{align}
&\Gamma(\Upsilon\to H_{c\bar{c}}+\gamma)\nonumber\\=&
\frac{m_{\Upsilon}^2-m^2_{H}}{16\pi
m_{\Upsilon}^3}|\overline{{\cal M}}(\Upsilon\to H_{c\bar{c}}+\gamma)|^2\nonumber\\=&\sum_{ij} C_{ij}(\mu)\langle 0|{\cal O}^{H_{c\bar{c}}}_i(\mu)|0\rangle \langle \Upsilon|{\cal O}_j(\mu)|\Upsilon\rangle,
\end{align}
where the factor $\langle \Upsilon|{\cal O}_j(\mu)|\Upsilon\rangle$ is  nonperturbative
matrix element for four-fermion annihilation operator in the bottomonium state, while the factor
$\langle 0|{\cal O}^{H_{c\bar{c}}}_i(\mu)|0\rangle$ is nonperturbative vacuum matrix element of
four-fermion production operator~\cite{Bodwin:1994jh}.
For a color-singlet operator, there is a simple
relation $\langle 0|{\cal O}^{H_{c\bar{c}}}_i(\mu)|0\rangle\approx (2J+1)\langle H_{c\bar{c}}|{\cal O}_i(\mu)|H_{c\bar{c}}\rangle$
with the spin quantum number $J$ of $H$ when consider the vacuum saturation approximation.

In our case, for the order of $\alpha_s^4 v^2$, the following local four-fermion operators
should be considered. For the $\Upsilon$ decay, we have the four-fermion annihilation operators
\begin{eqnarray}
\mathcal{O}(^{3}S_{1}^{[1]})&=&\psi^{\dagger}\bfsigma\chi\cdot\chi^{\dagger}\bfsigma\psi,\\
\mathcal{P}(^{3}S_{1}^{[1]})&=&\frac{1}{2}\left[\psi^{\dagger}\bfsigma
\chi\cdot\chi^{\dagger}\bfsigma(-\frac{i}{2}{\overleftrightarrow{ {\bold D}}})^2\psi+h.c.\right],\\
\mathcal{O}(^{1}S_{0}^{[8]})&=&\psi^{\dagger}T^a
\chi\cdot\chi^{\dagger}T^a\psi,\\
\mathcal{O}(^{3}S_{1}^{[8]})&=&\psi^{\dagger}T^a\bfsigma
\chi\cdot\chi^{\dagger}T^a\bfsigma\psi,
\end{eqnarray}
where $h.c.$ means the related complex conjugate term.

\begin{figure*}[t]
\includegraphics[width=0.3\textwidth]{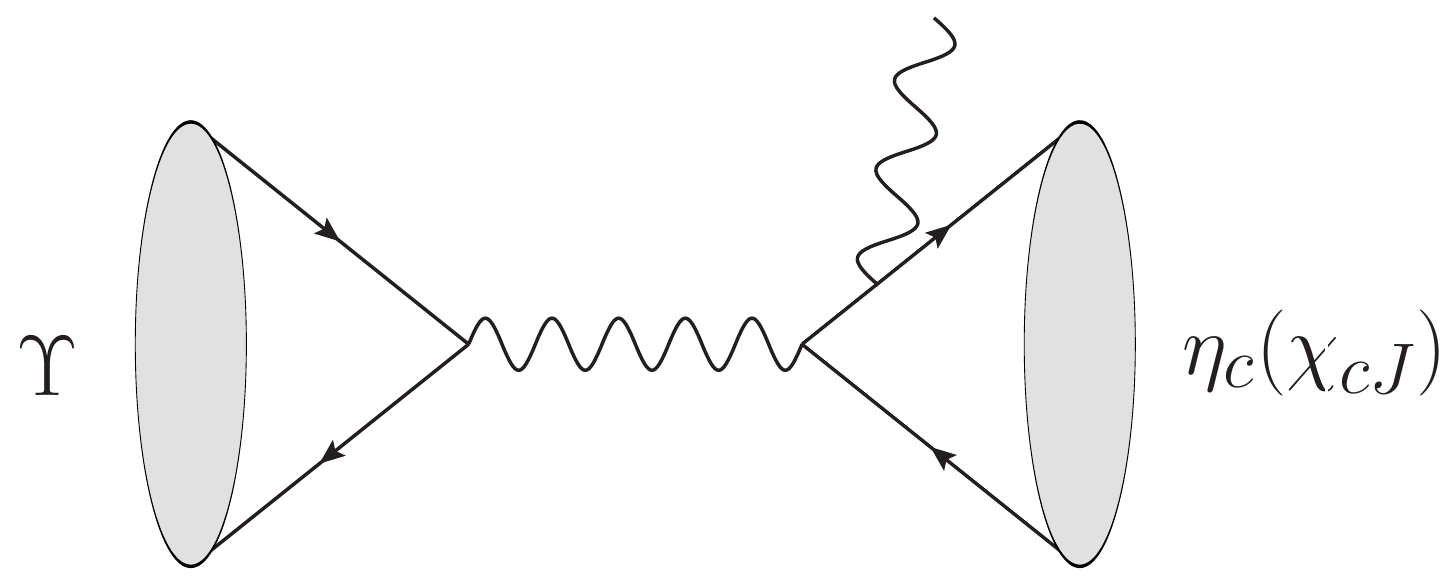}
\caption{
Typical QED Feynman diagrams for $\Upsilon$ radiative decays into $\eta_{c}$ or $\chi_{cJ}$.}
\label{QED}
\end{figure*}

\begin{figure*}[t]
\includegraphics[width=0.9\textwidth]{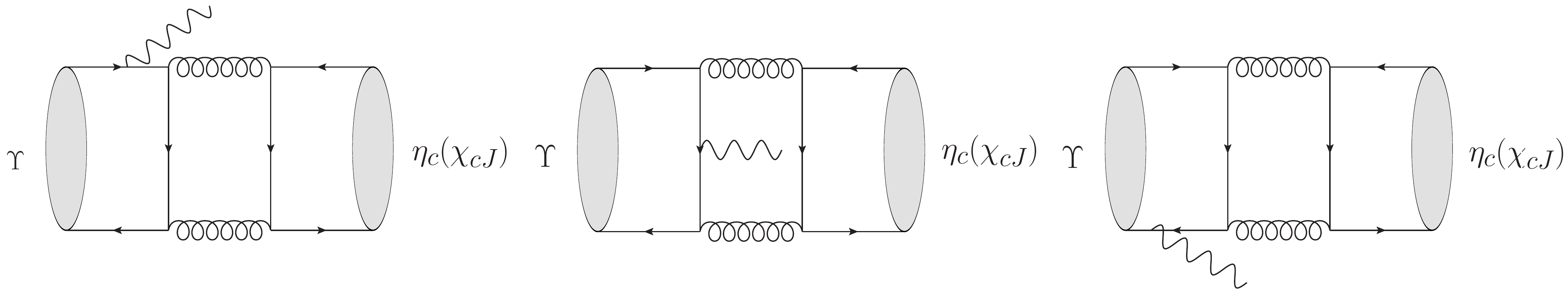}
\caption{
Typical QCD Feynman diagrams for $\Upsilon$ radiative decays into $\eta_{c}$ or $\chi_{cJ}$.}
\label{QCD}
\end{figure*}

For $\eta_c$ production, we have  the four-fermion production operators
\begin{eqnarray}
\mathcal{O}^H(^{1}S_{0}^{[1]})&=&\chi^{\dagger}\psi(a_H^\dagger a_H)\psi^{\dagger}\chi,\\
\mathcal{P}^H(^{1}S_{0}^{[1]})&=&\frac{1}{2}\left[\chi^{\dagger}\psi(a_H^\dagger a_H)\psi^{\dagger}(-\frac{i}{2}{\overleftrightarrow{ {\bold D}}})^2\chi+h.c.\right].~~~~~
\end{eqnarray}

For $P$-wave quarkonium production, we have
\begin{eqnarray}
\mathcal{O}^H(^{1}P_{0}^{[1]})&=&\frac{1}{3}\chi^{\dagger}(-\frac{i}{2}{\overleftrightarrow{ {\bold D}}}\cdot \bfsigma)\psi(a_H^\dagger a_H)\nonumber\\&&\times\psi^{\dagger}(-\frac{i}{2}{\overleftrightarrow{ {\bold D}}}\cdot \bfsigma)\chi,\\
\mathcal{O}^H(^{3}P_{1}^{[1]})&=&\frac{1}{2}\chi^{\dagger}(-\frac{i}{2}{\overleftrightarrow{ {\bold D}}}\times \bfsigma)\psi(a_H^\dagger a_H)\nonumber\\&&\times\psi^{\dagger}(-\frac{i}{2}{\overleftrightarrow{ {\bold D}}}\times \bfsigma)\chi,\\
\mathcal{O}^H(^{3}P_{2}^{[1]})&=&\chi^{\dagger}(-\frac{i}{2}{\overleftrightarrow{ {D}^i}}\sigma^j)\psi(a_H^\dagger a_H)\nonumber\\&&\times\psi^{\dagger}(-\frac{i}{2}{\overleftrightarrow{ { D}^i}} \sigma^j)\chi.
\end{eqnarray}

\subsection{Covariant projection method}

To calculate the amplitudes ${\cal M}$, we will apply the spin and color projection. For heavy quarkonium, we  define the momenta of quark and antiquark as
$p_Q=p_1=\frac{1}{2}P_H+k$ and $p_{\bar{Q}}=p_2=\frac{1}{2}P_H-k$. Then the spin projection is defined as

\begin{align}
&\Pi_{\Gamma_S}\left(p_{1}, p_{2}\right) \nonumber\\=&\sum_{s_{1}, s_{2}} u\left(p_{1}, s_{1}\right) \bar{v}\left(p_{2}, s_{2}\right)\left\langle\frac{1}{2}, s_{1} ; \frac{1}{2}, s_{2} \mid S S_z\right\rangle \nonumber\\
=&-\frac{\left({p}\!\!\!\slash_{1}+m_Q\right)(P\!\!\!\!\slash_H+2 E) \Gamma_S\left({p}\!\!\!\slash_{2}-m_Q\right)}{8 \sqrt{2} E^{2}(E+m_Q)},
\end{align}
where $E=\sqrt{m_Q^2+|\mathbf{k}|^2}$.  For the spin-singlet, we have the spin $S=0$ and $\Gamma_{S}=\gamma^5$. While for the spin-triplet, we have the spin $S=1$ and $\Gamma_{S=1}=\epsilon\!\!\!\slash_{H}=\epsilon_\mu(p_H) \gamma^\mu$.

The color projections are given as
\begin{align}
\mathcal{C}^{[1]}=\frac{\delta_{ij}}{\sqrt{N_c}},&~~~\mathcal{C}^{[8]}=\sqrt{2}T^a_{ij}.
\end{align}

The amplitudes defined in Eqs.~(\ref{LI}-\ref{LI-4}) can be obtained by combining the full standard model amplitudes ${\cal A}=\bar{v}{\cal A}_1 u \bar{u}{\cal A}_2 v$ and the projections
\begin{align}
&\mathcal{M}_{\mathcal{O}(^{3}S_{1}^{[a]}),\mathcal{O}^H(^{1}S_{0}^{[b]})}\nonumber\\=&\left. \mathrm{Tr}[\mathcal{C}^{[a]} \Pi_{1} \mathcal{A}_1] \mathrm{Tr}[\mathcal{C}^{[b]} \Pi^{\dag}_{0} \mathcal{A}_2]\right|_{k=0}, \\
&\mathcal{M}_{\mathcal{O}(^{3}S_{1}^{[a]}),\mathcal{P}^H(^{1}S_{0}^{[1]})}\nonumber\\=&\left. \frac{ {\cal R}'_{\mu\nu}\partial^{2} }{\partial k^{\mu} \partial k^{\nu}}\mathrm{Tr}[\mathcal{C}^{[a]} \Pi_{1} \mathcal{A}_1] \mathrm{Tr}[\mathcal{C}^{[1]} \Pi^{\dag}_{0} \mathcal{A}_2]\right|_{k=0}, \\
&\mathcal{M}_{\mathcal{O}(^{3}S_{1}^{[a]}),\mathcal{O}^H(^{3}P_{J}^{[1]})}\nonumber\\=&\left.
\frac{ \epsilon^{*(J)}_{\mu\nu}\partial }{\partial k^{\nu}}\mathrm{Tr}[\mathcal{C}^{[a]} \Pi_{1} \mathcal{A}_1] \mathrm{Tr}[\mathcal{C}^{[b]} \Pi^{\dag,\mu}_{1} \mathcal{A}_2]\right|_{k=0}(J=0,1,2),
\end{align}
where $ {\cal R}'_{\mu\nu}\partial^{2}$ is the projection for relativistic correction and its expression can be written as~\cite{Zhu:2017lwi,Zhu:2017lqu}
\begin{align}
{\cal R}'_{\mu\nu}=\frac{|\mathbf{k}|^{2}}{2(D-1)}\left(-g^{\mu \nu}+\frac{P_{H}^{\mu} P_{H}^{\nu}}{P_{H}^{2}}\right).
\end{align}
The sum over the polarization  for the P-wave charmonium are given in D-dimension~\cite{Petrelli:1997ge}

\begin{align}
\sum_{\mathrm{J_z}} \epsilon_{\alpha \beta}^{(0)} \epsilon_{\alpha^{\prime} \beta^{'}}^{*(0)} &=\frac{1}{D-1} \Pi_{\alpha \beta} \Pi_{\alpha^{\prime} \beta^{\prime}}, \\
\sum_{\mathrm{J_z}} \epsilon_{\alpha \beta}^{(1)} \epsilon_{\alpha^{\prime} \beta^{'}}^{*(1)} &=\frac{1}{2}\left(\Pi_{\alpha \alpha^{\prime}} \Pi_{\beta \beta^{\prime}}-\Pi_{\alpha \beta^{\prime}} \Pi_{\alpha^{\prime} \beta}\right), \\
\sum_{\mathrm{J_z}} \epsilon_{\alpha \beta}^{(2)} \epsilon_{\alpha^{\prime} \beta^{'}}^{*(2)} &=\frac{1}{2}\left(\Pi_{\alpha \alpha^{\prime}} \Pi_{\beta \beta^{\prime}}+\Pi_{\alpha \beta^{\prime}} \Pi_{\alpha^{\prime} \beta}\right)\nonumber\\&~~-\frac{1}{D-1} \Pi_{\alpha \beta} \Pi_{\alpha^{\prime} \beta^{\prime}},
\end{align}
where
\begin{align}
\Pi_{\alpha \alpha^{\prime}}&=\sum\epsilon_{\alpha} \epsilon_{\alpha^{\prime}}^{*} =-g^{\mu \nu}+\frac{P_{H}^{\mu} P_{H}^{\nu}}{P_{H}^{2}}.
\end{align}

\begin{figure*}[t]
\includegraphics[width=0.3\textwidth]{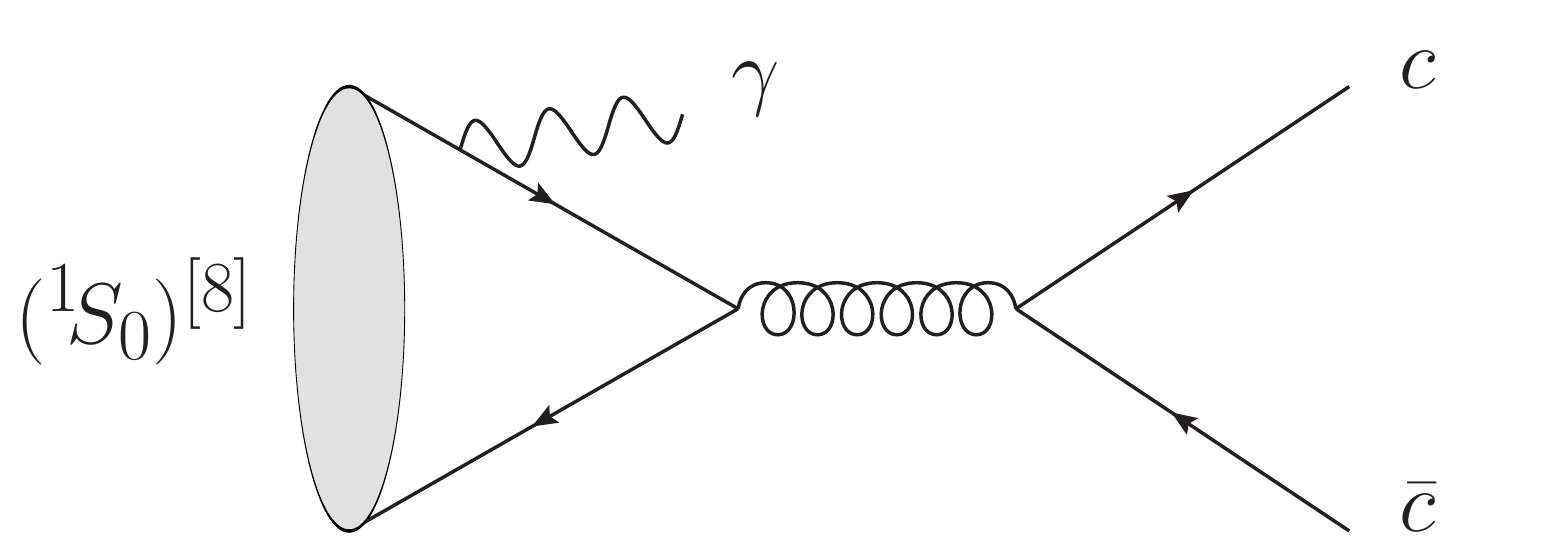}
\caption{
Feynman diagrams for color-octet $b\bar{b}$ into open charm by emitting a photon.}
\label{color}
\end{figure*}

\section{Color-singlet contribution}
The typical Feynman diagrams for color-singlet process are plotted in Figs.~\ref{QED} and \ref{QCD}. We have single-photon mediated mode in Fig.~\ref{QED}, which is pure QED diagram and at the order of $\alpha^3\sim 10^{-7}$. We also have $\gamma gg$ mediated mode in Fig.~\ref{QCD}, which are one-loop diagrams but at the order of $\alpha_s^4\alpha\sim 10^{-5}$.

Up to order of $v^2$, the color-singlet LDMEs contributes to  $\Upsilon\to\eta_c+\gamma$ are $\langle \Upsilon|\mathcal{O}(^{3}S_{1}^{[1]})|\Upsilon\rangle$, $\langle \Upsilon|\mathcal{P}(^{3}S_{1}^{[1]})|\Upsilon\rangle$, $\langle 0|\mathcal{O}^{\eta_c}(^{1}S_{0}^{[1]}))|0\rangle$
and $\langle 0|\mathcal{P}^{\eta_c}(^{1}S_{0}^{[1]}))|0\rangle$. Averaging over initial spin states we have the amplitude squared
\begin{align}
&|\overline{{\cal M}}|^2(\Upsilon(^{3}S_{1}^{[1]})\to\eta_c+\gamma)\nonumber\\=&\frac{1}{6}  \left(m_{\Upsilon }^2-m_{\eta _c}^2\right)^2\left[|a'|^2
\frac{\langle \Upsilon|\mathcal{O}(^{3}S_{1}^{[1]})|\Upsilon\rangle\langle 0|\mathcal{O}^{\eta_c}(^{1}S_{0}^{[1]}))|0\rangle}{5184m_{\Upsilon }m_{\eta _c}}
\right.\nonumber\\&\left.+(a_{v_2}' {a'}^{\dagger }+a' {a_{v_2}'}^{\dagger })
\frac{\langle \Upsilon|\mathcal{O}(^{3}S_{1}^{[1]})|\Upsilon\rangle\langle 0|\mathcal{P}^{\eta_c}(^{1}S_{0}^{[1]}))|0\rangle}{5184m_{\Upsilon }m_{\eta _c}}
\right.\nonumber\\&\left.+(a_{v_2}'' {a'}^{\dagger }+a' {a_{v_2}''}^{\dagger })
\frac{\langle \Upsilon|\mathcal{P}(^{3}S_{1}^{[1]})|\Upsilon\rangle\langle 0|\mathcal{O}^{\eta_c}(^{1}S_{0}^{[1]}))|0\rangle}{5184m_{\Upsilon }m_{\eta _c}}\right],
\end{align}
where $a'$, $a_{v_2}'$, $a_{v_2}''$ and in the following $b', c', d', e', f', g', h', x'$ and $y'$ are also Lorentz invariants and only include the short-distance interactions and the nonperturbative matrix elements with additional factors are removed compared with the parameters $a, b, c, d, e, f, g, h, x$, and $y$ defined in Eqs.~(\ref{LI}-\ref{LI-4}).

At order of $v^2$, we only need to consider the color-singlet LDMEs $\langle \Upsilon|\mathcal{O}(^{3}S_{1}^{[1]})|\Upsilon\rangle$ and $\langle 0|\mathcal{O}^{\chi_{cJ}}(^{1}S_{0}^{[1]}))|0\rangle$ in $\Upsilon\to\chi_{cJ}+\gamma$. Averaging over initial spin states we have the following amplitude squared
\begin{align}
&|\overline{{\cal M}}|^2(\Upsilon(^{3}S_{1}^{[1]})\to\chi_{c0}+\gamma)\nonumber\\=&\frac{\langle \Upsilon|\mathcal{O}(^{3}S_{1}^{[1]})|\Upsilon\rangle\langle 0|\mathcal{O}^{\chi_{c0}}(^{3}P_{0}^{[1]}))|0\rangle}{5184m_{\Upsilon }m_{\chi_{c0}} }[|b'|^2+\frac{1}{6} \left(b' {c'}^{\dagger }\right.\nonumber\\&\left.+c' {b'}^{\dagger } +\frac{1}{2} |c'|^2 \left(m_{\Upsilon }^2-m_{\chi _{{c0}}}^2\right)\left(m_{\Upsilon }^2-m_{\chi_{{c0}}}^2\right)\right)],
\end{align}

\begin{align}
&|\overline{{\cal M}}|^2(\Upsilon(^{3}S_{1}^{[1]})\to\chi_{c1}+\gamma)\nonumber\\=&\frac{\langle \Upsilon|\mathcal{O}(^{3}S_{1}^{[1]})|\Upsilon\rangle\langle 0|\mathcal{O}^{\chi_{c1}}(^{3}P_{1}^{[1]}))|0\rangle}{15552m_{\Upsilon }m_{\chi_{c1}}}[
\frac{|e'|^2 \left(m_{\Upsilon }^2-m_{\chi
   _{c1}}^2\right)^4}{24 m_{\Upsilon }^2}\nonumber\\&+\frac{|d'|^2
   \left(m_{\Upsilon }^2+ m_{\chi _{c1}}^2\right)\left(5m_{\Upsilon }^2+m_{\chi _{c1}}^2\right)}{6 m_{\Upsilon }^2}+\frac{\left(m_{\Upsilon }^2-m_{\chi _{c1}}^2\right)^2}{6}\nonumber\\&\times\left(|f'|^2 m_{\Upsilon }^2 -\frac{({d'}^{\dagger } e'+{e'}^{\dagger } d') \left(m_{\Upsilon }^2+m_{\chi _{c1}}^2\right) }{2 m_{\Upsilon }^2}\right.\nonumber\\&\left.-
  \left({d'}^{\dagger } f'+{f'}^{\dagger } d'\right)\frac{2m_{\Upsilon }^2}{m_{\Upsilon }^2-m_{\chi _{c1}}^2}\right)],
\end{align}

\begin{align}
&|\overline{{\cal M}}|^2(\Upsilon(^{3}S_{1}^{[1]})\to\chi_{c2}+\gamma)\nonumber\\=&\frac{\langle \Upsilon|\mathcal{O}(^{3}S_{1}^{[1]})|\Upsilon\rangle\langle 0|\mathcal{O}^{\chi_{c2}}(^{3}P_{2}^{[1]}))|0\rangle}{25920m_{\Upsilon }m_{\chi_{c2}}}[
\frac{\left(m_{\Upsilon }^2-m_{\chi _{c2}}^2\right){}^4}{24 m_{\chi _{c2}}^4}(|y'|^2\nonumber\\&+\frac{5m_{\chi _{c2}}^2}{6m_{\Upsilon }^2 }|x'|^2 -\frac{g'{y'}^\dag+y'{g'}^\dag}{3m_{\Upsilon }^2 })+(x'{y'}^\dag+y'{x'}^\dag+h'{x'}^\dag\nonumber\\&
{\small+x'{h'}^\dag-\frac{5 m_{\chi _2}^2(g'{x'}^\dag+x'{g'}^\dag)}{3\left(m_{\Upsilon }^2-m_{\chi _2}^2\right)^2})\frac{\left(m_{\Upsilon }^2-m_{\chi _2}^2\right)^4 \left(m_{\Upsilon }^2+m_{\chi _2}^2\right)}{144 m_{\Upsilon }^2 m_{\chi _2}^4}}\nonumber\\&+\frac{5 |g|^2 \left(10 m_{\Upsilon }^2 m_{\chi _{c2}}^2+m_{\Upsilon }^4+m_{\chi _{c2}}^4\right)}{36 m_{\Upsilon }^2 m_{\chi _{c2}}^2}-(\frac{x'{y'}^\dag+y'{x'}^\dag}{m_{\Upsilon }^2}\nonumber\\&+ g'{h'}^\dag+h'{g'}^\dag) \frac{\left(m_{\Upsilon }^2-m_{\chi _2}^2\right){}^2 \left(8 m_{\Upsilon }^2 m_{\chi _2}^2+m_{\Upsilon }^4+m_{\chi _2}^4\right)}{72 m_{\Upsilon }^2
   m_{\chi _2}^4}].
\end{align}
In total, the analytic expression for $a'$, $a_{v_2}'$, $a_{v_2}''$, $b', c', d', e', f', g', h', x'$ and $y'$ are given in the appendix.

\section{Color-octet contribution}
In general, the evidence of color-octet contribution exists only in the inclusive processes~\cite{Bodwin:1994jh,Chao:2012iv}.
In this section, we will discuss the possibility of color-octet contribution in
the radiative exclusive decays of upsilon into charmonium.

For simplicity, we divide the color-octet contribution into two steps. First the upsilon decays into a pair
of open-charm and a photon, and then the open-charm can be fragmented into a charmonium. The Feynman diagrams for the upsilon decays into a pair
of open-charm and a photon are plotted in Fig.~\ref{color}. Thus the
color-octet contribution to the decay width can be written as
\begin{align}
&\Gamma(\Upsilon(^{1}S_{0}^{[8]}, ~^{3}S_{1}^{[8]})\to H_{c\bar{c}}(p)+\gamma+X)\nonumber\\=&
\sum_i \int_0^1 d z \Gamma(\Upsilon(^{1}S_{0}^{[8]}, ^{3}S_{1}^{[8]})\to i(p/z)+\gamma)D_{i\to H_{c\bar{c}}}(z)  ,
\end{align}
where the $D_{i\to H_{c\bar{c}}}(z)$ is the fragmentation function.
Keeping only the charm quark and antiquark contributions, the leading order gives
\begin{align}
&\Gamma(\Upsilon(^{1}S_{0}^{[8]}, ^{3}S_{1}^{[8]})\to H_{c\bar{c}}(p)+\gamma+X)\nonumber\\=&2\Gamma(\Upsilon(^{1}S_{0}^{[8]}, ^{3}S_{1}^{[8]})\to i(p/z)+\gamma)\int_0^1 d z D_{i\to H_{c\bar{c}}}(z,\mu)  ,
\end{align}
The fragmentation functions for production of S-wave and P-wave heavy quarkonium can be found in Refs.~\cite{Braaten:1993mp,Yuan:1994hn,Qiao:2011yk}. In the appendix, we will list the fragmentation probabilities.

\section{Phenomenological discussions}
In the numerical calculation, we need to input  the heavy quark mass, the heavy quark relativistic velocity, the coupling constants and the LDMEs. For the first two parameters, we just input heavy quark relativistic velocity and extract the heavy quark mass with the on-shell condition. We first take the bottom relative velocity squared as $\left|\mathbf{v}\right|^2\simeq0.1$ in bottomonium and the charm relative velocity squared as $\left|\mathbf{v}\right|^2\simeq0.3$ in charmonium~\cite{Bodwin:1994jh}. For heavy quarkonium, we have $m_{H}=2\sqrt{m_{Q}^2+\mathbf{k}^{2}}\simeq 2m_{Q}+\frac{\mathbf{k^{2}}}{2m_{Q}}= 2m_{Q}(1+\frac{\mathbf{v^{2}}}{16})$ with $\left|\mathbf{k}\right|^2=m_{Q}^2 \left|\mathbf{v}\right|^2/4$. For $\Upsilon\to \eta_{c}+\gamma$, we have $m_{\Upsilon}=9.4603\mathrm{GeV}, m_{\eta_{c}}=2.984\mathrm{GeV}$~\cite{Tanabashi:2018oca}. Then we can extract
the quark masses as $m_b=4.67\mathrm{GeV}, m_c=1.44\mathrm{GeV}$.  For the case of $\Upsilon\to \chi_{cJ}+\gamma$, the corresponding quarkonium masses are: $m_{\chi_{c0}}=3.415\mathrm{GeV}, m_{\chi_{c1}}=3.511\mathrm{GeV}, m_{\chi_{c2}}=3.566\mathrm{GeV}$~\cite{Tanabashi:2018oca}, we can extract the charm quark mass
as  $m_c=1.56\mathrm{GeV}$ for P-wave charmonium. These choices of heavy quark masses are also consistent with previous
literatures~\cite{Qiao:2012vt,Qiao:2012hp,Qiao:2014pfa,Zhu:2016udl}. The coupling constant are chosen as: $\alpha=1/128, \alpha_s(2m_b)=0.182, \alpha_s(m_b)=0.219, \alpha_s(m_c)=0.345$. To calculate the one-loop integrals, we use Package-X in Ref.~\cite{Patel:2015tea}.

The NRQCD LDMEs are processes independent thus we can extract them from the heavy quarkonium leptonic decays or potential models. For example, the leptonic decay width is given as
 \begin{align}
\Gamma(\Upsilon\to \mu^+\mu^-)=&
 \frac{16\pi e_b^2\alpha^2}{ m_{\Upsilon }^2}(1-\frac{4\alpha_s C_F}{\pi}) \frac{\langle \Upsilon|\mathcal{O}(^{3}S_{1}^{[1]})|\Upsilon\rangle}{2N_c}.
\end{align}

The color-singlet LDMEs are determined as: $\langle \Upsilon|\mathcal{O}(^{3}S_{1}^{[1]})|\Upsilon\rangle=2.925GeV^3$, $\langle 0|\mathcal{O}^{\eta_c}(^{1}S_{0}^{[1]}))|0\rangle=0.41GeV^3$,  $\langle \Upsilon|\mathcal{P}(^{3}S_{1}^{[1]})|\Upsilon\rangle\approx |\mathbf{k}|^2\langle \Upsilon|\mathcal{O}(^{3}S_{1}^{[1]})|\Upsilon\rangle=0.34GeV^5$, $\langle 0|\mathcal{P}^{\eta_c}(^{1}S_{0}^{[1]}))|0\rangle\approx |\mathbf{k}|^2\langle 0|\mathcal{O}^{\eta_c}(^{1}S_{0}^{[1]}))|0\rangle=0.044GeV^5$~\cite{Zhu:2017lwi,Zhu:2017lqu}.
The heavy quark spin symmetry is hold and we have $\langle 0|\mathcal{O}^{\chi_{c2}}(^{3}P_{2}^{[1]}))|0\rangle/5=\langle 0|\mathcal{O}^{\chi_{c1}}(^{3}P_{1}^{[1]}))|0\rangle/3=\langle 0|\mathcal{O}^{\chi_{c0}}(^{3}P_{0}^{[1]}))|0\rangle=0.107GeV^5$~\cite{Zhu:2017lwi,Zhu:2017lqu}.

For the color-octet LDMEs, their values are suppressed for $\Upsilon$ decays.
In Ref.~\cite{Bodwin:2005gg}, one can find that $\langle \Upsilon|\mathcal{O}(^{1}S_{0}^{[8]})|\Upsilon\rangle/\langle \Upsilon|\mathcal{O}(^{3}S_{1}^{[1]})|\Upsilon\rangle=2.414\times 10^{-3}$
and $\langle \Upsilon|\mathcal{O}(^{3}S_{1}^{[8]})|\Upsilon\rangle/\langle \Upsilon|\mathcal{O}(^{3}S_{1}^{[1]})|\Upsilon\rangle=8.1\times 10^{-5}$. Thus we only consider the $\mathcal{O}(^{1}S_{0}^{[8]})$ contributions in $\Upsilon$ decays.

Using the previous formalism, we can easily get the branching fraction
\begin{align}
{\cal B}(\Upsilon\to H_{c\bar{c}}+\gamma)=&
\frac{\Gamma(\Upsilon\to H_{c\bar{c}}+\gamma)}{\Gamma(\Upsilon)},
\end{align}
where $\Gamma(\Upsilon)$ is the total decay width. From PDG, one can get $\Gamma(\Upsilon(1S))=54.02$keV, $\Gamma(\Upsilon(2S))=31.98$keV, and $\Gamma(\Upsilon(1S))=20.32$keV~\cite{Tanabashi:2018oca}. However, to reduce the theoretical uncertainty, one can use the results for the $\Upsilon\to \mu^+\mu^-$  or $\Upsilon\to ggg$ and then the uncertainty from LDMEs of $\Upsilon(nS)$ can be eliminated.  The decay width for  $\Upsilon\to ggg$ is~\cite{Schuler:1994hy}
 \begin{align}
\Gamma(\Upsilon\to ggg)=&
 \frac{80 \left(\pi ^2-9\right) \alpha _s^3}{243 m_{\Upsilon }^2}\langle \Upsilon|\mathcal{O}(^{3}S_{1}^{[1]})|\Upsilon\rangle
 ,
\end{align}
In this process, we may vary the renormalization scale from $\mu=m_c$ to $\mu=2m_b$. In PDG, one can find that $\Gamma(\Upsilon\to ggg)/\Gamma(\Upsilon)=(81.7\pm0.7)\%$~\cite{Tanabashi:2018oca}.  We then give the numerical results of the branching fractions for the upsilon radiative decays into a charmonium in Tabs.~\ref{table1}, ~\ref{table2}, ~\ref{table3} and \ref{table4}. For other channels with radially excited heavy quarkonium, we also give the predictions of the branching fractions in Tab.~\ref{table5}.

\begin{table}[h]
\begin {center}
\begin{tabular}{c|c|c|c}
 \hline\hline
Contribution&$\mu=m_c$&$\mu=m_b$
&$\mu=2m_b$\\\hline
${\cal B}_{QED}$&$1.93\times 10^{-6}$&$1.93\times 10^{-6}$&$1.93\times 10^{-6}$\\
${\cal B}_{QCD}$&$2.93\times 10^{-5}$&$2.82\times 10^{-5}$&$2.58\times 10^{-5}$\\
${\cal B}_{Rel.Cor.}$&$-4.62\times 10^{-6}$&$-4.45\times 10^{-6}$&$-4.07\times 10^{-6}$\\
${\cal B}_{Col.Oct.}$&$2.79\times 10^{-8}$&$2.69\times 10^{-8}$&$2.46\times 10^{-8}$\\
${\cal B}_{In~total}$&$2.80\times 10^{-5}$&$3.19\times 10^{-5}$&$3.29\times 10^{-5}$\\
\hline\hline
\end{tabular}
\caption{Branching fractions for the process of
$\Upsilon\to\eta_{c}+\gamma$ with different renormalization scale $\mu$. ${\cal B}_{QED}$ represents
the pure QED contribution; ${\cal B}_{QCD}$ represents
the $\gamma gg$ (QCD) contribution; ${\cal B}_{Rel.Cor.}$ represents
the relativistic correction; ${\cal B}_{Col.Oct.}$ represents
the color-octet contribution; ${\cal B}_{In~total}$ represents the total predictions.  }
 \label{table1}
\end {center}
\vspace{-0.5cm}
\end{table}

\begin{table}[h]
\begin {center}
\begin{tabular}{c|c|c|c}
 \hline\hline
Contribution&$\mu=m_c$&$\mu=m_b$
&$\mu=2m_b$\\\hline
${\cal B}_{QED}$&$2.56\times 10^{-8}$&$2.56\times 10^{-8}$&$2.56\times 10^{-8}$\\
${\cal B}_{QCD}$&$1.76\times 10^{-6}$&$1.66\times 10^{-6}$&$1.51\times 10^{-6}$\\
${\cal B}_{Col.Oct.}$&$2.18\times 10^{-9}$&$2.05\times 10^{-9}$&$1.88\times 10^{-9}$\\
${\cal B}_{In~total}$&$1.82\times 10^{-6}$&$1.80\times 10^{-6}$&$1.72\times 10^{-6}$\\
\hline\hline
\end{tabular}
\caption{Branching fractions for the process of
$\Upsilon\to\chi_{c0}+\gamma$ with different renormalization scale $\mu$.  }
 \label{table2}
\end {center}
\vspace{-0.5cm}
\end{table}

\begin{table}[h]
\begin {center}
\begin{tabular}{c|c|c|c}
 \hline\hline
Contribution&$\mu=m_c$&$\mu=m_b$
&$\mu=2m_b$\\\hline
${\cal B}_{QED}$&$4.85\times 10^{-7}$&$4.85\times 10^{-7}$&$4.85\times 10^{-7}$\\
${\cal B}_{QCD}$&$9.22\times 10^{-7}$&$8.70\times 10^{-7}$&$7.96\times 10^{-7}$\\
${\cal B}_{Col.Oct.}$&$2.51\times 10^{-9}$&$2.37\times 10^{-9}$&$2.16\times 10^{-9}$\\
${\cal B}_{In~total}$&$1.61\times 10^{-6}$&$2.54\times 10^{-6}$&$3.29\times 10^{-6}$\\
\hline\hline
\end{tabular}
\caption{Branching fractions for the process of
$\Upsilon\to\chi_{c1}+\gamma$ with different renormalization scale $\mu$.  }
 \label{table3}
\end {center}
\vspace{-0.5cm}
\end{table}

\begin{table}[h]
\begin {center}
\begin{tabular}{c|c|c|c}
 \hline\hline
Contribution&$\mu=m_c$&$\mu=m_b$
&$\mu=2m_b$\\\hline
${\cal B}_{QED}$&$1.49\times 10^{-7}$&$1.49\times 10^{-7}$&$1.49\times 10^{-7}$\\
${\cal B}_{QCD}$&$4.95\times 10^{-6}$&$4.67\times 10^{-6}$&$4.27\times 10^{-6}$\\
${\cal B}_{Col.Oct.}$&$9.47\times 10^{-10}$&$8.93\times 10^{-10}$&$8.17\times 10^{-10}$\\
${\cal B}_{In~total}$&$5.45\times 10^{-6}$&$5.76\times 10^{-6}$&$5.80\times 10^{-6}$\\
\hline\hline
\end{tabular}
\caption{Branching fractions for the process of
$\Upsilon\to\chi_{c2}+\gamma$ with different renormalization scale $\mu$.  }
 \label{table4}
\end {center}
\vspace{-0.5cm}
\end{table}

\begin{table}[h]
\begin {center}
\begin{tabular}{c|c}
 \hline\hline
Channels&${\cal B}(\times10^{-5})$\\\hline
$\Upsilon\to \eta_c+\gamma$&$3.19_{-0.39}^{+0.10}$\\
$\Upsilon(2S)\to \eta_c+\gamma$&$4.23_{-0.52}^{+0.13}$\\
$\Upsilon(3S)\to \eta_c+\gamma$&$4.83_{-0.59}^{+0.15}$\\
$\Upsilon\to \eta_c(2S)+\gamma$&$1.34_{-0.17}^{+0.04}$\\
$\Upsilon\to \chi_{c0}+\gamma$&$0.18_{-0.01}^{+0.01}$\\
$\Upsilon\to \chi_{c1}+\gamma$&$0.25_{-0.09}^{+0.08}$\\
$\Upsilon\to \chi_{c2}+\gamma$&$0.58_{-0.03}^{+0.01}$\\
$\Upsilon(2S)\to \chi_{c0}+\gamma$&$0.21_{-0.01}^{+0.01}$\\
$\Upsilon(2S)\to \chi_{c1}+\gamma$&$0.28_{-0.10}^{+0.08}$\\
$\Upsilon(2S)\to \chi_{c2}+\gamma$&$0.92_{-0.05}^{+0.01}$\\
$\Upsilon(3S)\to \chi_{c0}+\gamma$&$0.22_{-0.01}^{+0.01}$\\
$\Upsilon(3S)\to \chi_{c1}+\gamma$&$0.33_{-0.12}^{+0.10}$\\
$\Upsilon(3S)\to \chi_{c2}+\gamma$&$0.12_{-0.01}^{+0.01}$\\
\hline\hline
\end{tabular}
\caption{Branching fractions for the process of
$\Upsilon(nS)\to H_{c\bar{c}}+\gamma$. Here we give the
total prediction and the uncertainty is from renormalization scale $\mu\subset [m_c, 2m_b]$.  }
 \label{table5}
\end {center}
\vspace{-0.5cm}
\end{table}

The largest theoretical uncertainties are from the renormalization scale and LDMEs,  thus we choose the prediction at $\mu=m_b$ as a central value and vary the renormalization scale from $\mu=m_c$ to $\mu=2m_b$ for the former uncertainty. But these uncertainties can be reduced by considering the results for both the $\Upsilon\to \mu^+\mu^-$  and $\Upsilon\to ggg$ channels. From Tabs.~\ref{table1}, ~\ref{table2}, ~\ref{table3} and \ref{table4}, one can see the contributions from QED, color-octet mechanism, relativistic corrections are small compared the QCD contribution.
The theoretical predictions for the branching fractions of $\Upsilon(1S) \to \chi_{c0}+\gamma$ and $\Upsilon(1S) \to \chi_{c2}+\gamma$ are below the upper limits at Belle experiment, and the theoretical predictions for $\Upsilon(1S) \to \eta_{c}+\gamma$ is close to the upper limit at Belle experiment. However, the theoretical predictions for the branching fraction of $\Upsilon(1S) \to \chi_{c1}+\gamma$
is  about one order of magnitude smaller than the Belle data. Considering the experimental uncertainty is also large, more precise
theoretical calculation and experimental determination are required to understand the discrepancy in $\Upsilon(1S) \to \chi_{c1}+\gamma$ channel.

\section{Conclusion}

In this paper, we have calculated the branching fractions of the radiative decays of bottomonium into S-wave and P-wave charmonium within NRQCD approach. The analytical expression for these branching fractions are obtained. The relativistic corrections for the channel $\Upsilon(nS)\to \eta_c(nS)+\gamma$ are also obtained. Even though  both the color-singlet and color-octet contributions are included,
the theoretical prediction of the branching fraction for $\Upsilon(1S) \to \chi_{c1}+\gamma$ is  about one order of magnitude smaller than the very recent Belle data, which indicates the $\chi_{c1}$ production mechanism in upsilon decay is not well understood. Further studies from both the theoretical and experimental aspects are required to solve the $\Upsilon(1S) \to \chi_{c1}+\gamma$ puzzle.

\acknowledgments
The authors thank the useful discussions with Xiangpeng Wang. This work is supported by NSFC under grant No.~11705092, 11975127, 12075124, and by Jiangsu Qing Lan Project and  Jiangsu Specially Appointed Professor Program.
\begin{widetext}
\section*{Appendix}
In this appendix, we give the analytical expression for the Lorentz invariants in the decay widths and  the charmonium fragmentation probabilities.
For pure QED contribution, we have
\begin{align}
&a'=-\frac{32e_l^3 \mathit{N_{c}}^2}{27m^3_{b}(r^2-1)}, a_{v_2}'=\frac{64 e_l^3  \mathit{N_c}^2 (7r^2-2)}{81m_{b}^5 r^2(r^2-1)^2}, a_{v_2}''=\frac{64 e_l^3 \mathit{N_c}^2 (7r^2-2)}{81m_{b}^5 r^2(r^2-1)^2},
\end{align}

\begin{align}
&b'=-\frac{64 e_l^3 \mathit{N_{c}}^2(3 r^2-1)}{27 \sqrt{3} m_{b}^2 r(r^2-1)}, c'=-\frac{32 e_l^3 \mathit{N_{c}}^2 (3 r^2-1)}{27 \sqrt{3} m_{b}^4 r (r^2-1)^2},
\end{align}

\begin{align}
d'=-\frac{32 \sqrt{2} e_l^3 \mathit{N_{c}}^2}{27 m_{b}^3 r^2 (r^2-1)}, e'=\frac{16 \sqrt{2} e_l^3 \mathit{N_{c}}^2}{27 m_{b}^5 r^2 (r^2-1)^2}, f'=\frac{16 \sqrt{2} e_l^3 \mathrm{N_{c}}^2}{27 m_{b}^5 r^2 (r^2-1)^2},
\end{align}

\begin{align}
g'=\frac{128 e_l^3 \mathit{N_{c}}^2 r}{27 m_{b}^2 (r^2-1)}, h'=\frac{64 e_l^3 \mathit{N_{c}}^2 r}{27 m_{b}^4 (r^2-1)^2}, x'=\frac{64 e_l^3 \mathit{N_{c}}^2 r}{27 m_{b}^4 (r^2-1)^2}, y'=-\frac{64 e_l^3 \mathit{N_{c}}^2 r}{27 m_{b}^4 (r^2-1)^2},
\end{align}
where $r=\frac{m_c}{m_b}$, $e_l=\sqrt{4\pi\alpha}$  with $\alpha=1/128$.
For QCD contribution, we have
\begin{align}
a'=\alpha_{s}^2 e_l C_A C_F [-\frac{16 \mathit{B}_1}{3 m_{b}^3 \left(1-r^2\right)^2}+\frac{16 \mathit{B}_2}{3 m_{b}^3 \left(1-r^2\right)^2}-\frac{64
   \mathit{C}_1}{3 m_{b} \left(1-r^4\right)}-\frac{64 \mathit{C}_6 r^2}{3 m_{b} \left(1-r^4\right) }-\frac{32}{3 m_{b}^3
   \left(1-r^2\right)^2}],
\end{align}

\begin{align}
a_{v_2}'=\alpha_{s}^2e_lC_A C_F&[-\frac{16    \mathit{B}_1  (r^6-r^4+11 r^2+1)  }{9 m_{b}^5 r^2 (1-r^2)^4 (r^2+1)}-\frac{32  \mathit{B}_2  (-r^4-9 r^2+1)  }{9 m_{b}^5 r^2 (1-r^2)^3 (r^2+1)}-\frac{16   \mathit{B}_3 (-r^4-2 r^2+1)  }{3 m_{b}^5 r^2 (1-r^2)^4}\nonumber\\&+\frac{32   \mathit{B}_4   }{3 m_{b}^5 r^2 (1-r^2)^2(r^2+1)}+\frac{128   \mathit{C}_1  (-6 r^4-2 r^2+1)  }{9 m_{b}^3 r^2 (1-r^2)^2(r^2+1)^2}+\frac{32   \mathit{C}_2  (2-r^2)  }{9 m_{b}^3 r^2 (1-r^2)^2}\nonumber\\&-\frac{32 \mathit{C}_3 (2-5 r^2) }{9 m_{b}^3 r^2 (1-r^2)^2}+\frac{64   \mathit{C}_4  (1-7 r^4)  }{9
   m_{b}^3 (1-r^2)^2 (r^2+1)^2}-\frac{16    (7 r^4+30 r^2+11)  }{9 m_{b}^5 r^2 (1-r^2)^3(r^2+1)}],
\end{align}

\begin{align}
a_{v_2}''=\alpha_{s}^2e_lC_A C_F&[-\frac{8  \mathit{B}_1  (12 r^{12}+31 r^8-44 r^6+94 r^4-84 r^2+23) }{9 m_{b}^5 (1-2 r^2)^2 (1-r^2)^4
   (r^2+1)^2}+\frac{8  \mathit{B}_2  (-92 r^{10}-188 r^8-15 r^6+305 r^4-177 r^2+31) }{9 m_{b}^5 (1-2 r^2)^2
   (1-r^2)^3 (r^2+1)^2}\nonumber\\&+\frac{64  \mathit{B}_3  (-2 r^4+r^2+2) }{9 m_{b}^5 (1-r^2)^4}-\frac{32
    \mathit{B}_4  (r^2+3) (-r^4-3 r^2+2) }{9 m_{b}^5 (1-r^2)^3 (r^2+1)^2}+\frac{32
   \mathit{C}_1  (3 r^6+11 r^4+9 r^2+17) }{9 m_{b}^3 (1-r^2)^2 (r^2+1)^3}\nonumber\\&-\frac{16  \mathit{C}_2
   (1-7 r^2) }{3 m_{b}^3 (1-r^2)^2}-\frac{16  \mathit{C}_3  (r^2+9) }{9 m_{b}^3
   (1-r^2)^2}-\frac{32  \mathit{C}_4  (-r^{10}-2 r^6-8 r^4-29 r^2+8) }{9 m_{b}^3 (1-r^2)^3
   (r^2+1)^3}\nonumber\\&-\frac{16   (-74 r^{10}-95 r^8+17 r^6+115 r^4-59 r^2+8) }{9 m_{b}^5 (1-2 r^2)^2
   (1-r^2)^3 (r^2+1)^2}],
\end{align}

\begin{align}
b'=\alpha_{s}^2e_l C_A C_F&[-\frac{32  \mathit{B}_1  (r^4+r^2+4)}{3 \sqrt{3} m_{b}^2 r (r^2-1)^3}+\frac{32  \mathit{B}_2
   (r^4-9 r^2+4)}{3 \sqrt{3} m_{b}^2 r (r^2-1)^3}+\frac{320  \mathit{B}_3  r}{3 \sqrt{3} m_{b}^2
   (r^2-1)^3}+\frac{64  \mathit{C}_1  (5 r^4-16 r^2+3)}{3 \sqrt{3} (r^2-1)^2 (r^3+r)}\nonumber\\&-\frac{64
    \mathit{C}_2 }{\sqrt{3} r}-\frac{64  \mathit{C}_3  (r^2+1)}{3 \sqrt{3} r
   (r^2-1)}+\frac{128  \mathit{C}_5  r}{\sqrt{3} (r^2+1)}+\frac{32   (r^4+14 r^2-7) }{3 \sqrt{3} m_{b}^2 r (r^2-1)^3}],
\end{align}
\begin{align}
c'=\alpha_{s}^2e_lC_A C_F&[-\frac{16  \mathit{B}_1  (r^4+r^2+4) }{3 \sqrt{3} m_{b}^4 r (r^2-1)^4}+\frac{16  \mathit{B}_2
   (r^4-9 r^2+4) }{3 \sqrt{3} m_{b}^4 r (r^2-1)^4}+\frac{160  \mathit{B}_3  r }{3 \sqrt{3} m_{b}^4
   (r^2-1)^4}+\frac{32  \mathit{C}_1  (5 r^4-16 r^2+3) }{3 \sqrt{3} m_{b}^2 (r^2-1)^3
   (r^3+r)}\nonumber\\&-\frac{32  \mathit{C}_2  (2 r^4-5 r^2+1) }{3 \sqrt{3} m_{b}^2 r (r^2-1)^3}-\frac{32
    \mathit{C}_3  (2 r^4-r^2+1) }{3 \sqrt{3} m_{b}^2 r (r^2-1)^3}+\frac{64  \mathit{C}_5  r C_A
   C_F}{\sqrt{3} m_{b}^2 (r^4-1)}+\frac{16   (r^4+14 r^2-7) }{3 \sqrt{3} m_{b}^4 r (r^2-1)^4}],
\end{align}

\begin{align}
d'=\alpha_{s}^2e_lC_A C_F&[-\frac{16 \sqrt{2}  \mathit{B}_1  (r^4+2 r^2-1) }{3 m_{b}^3 r^2 (r^2-1)^3}-\frac{16 \sqrt{2}  \mathit{B}_2
    (7 r^4-6 r^2+1) }{3 m_{b}^3 r^2 (r^2-1)^3}+\frac{64 \sqrt{2}  \mathit{B}_3  (2 r^2-1) }{3
   m_{b}^3 (r^2-1)^3}-\frac{64 \sqrt{2}  \mathit{C}_1  }{3 m_{b} r^2 (r^2-1)^2}\nonumber\\&-\frac{32 \sqrt{2}
   \mathit{C}_2  (5 r^2-1) }{3 m_{b} r^2 (r^2-1)}-\frac{32 \sqrt{2}  \mathit{C}_3  (r^2+1) }{3
   m_{b} r^2 (r^2-1)}+\frac{32 \sqrt{2}   (7 r^4-6 r^2+1) }{3 m_{b}^3 r^2 (r^2-1)^3}],
\end{align}
\begin{align}
e'=\alpha_{s}^2e_lC_A C_F&[-\frac{8 \sqrt{2}  \mathit{B}_1  (r^4+4 r^2+1) }{3 m_{b}^5 r^2 (r^2-1)^4}+\frac{8 \sqrt{2}  \mathit{B}_2
    (-15 r^4+4 r^2+1) }{3 m_{b}^5 r^2 (r^2-1)^4}+\frac{128 \sqrt{2}  \mathit{B}_3  r^2 }{3 m_{b}^5
   (r^2-1)^4}+\frac{32 \sqrt{2}  \mathit{C}_1  }{3 m_{b}^3 r^2 (r^2-1)^3}\nonumber\\&-\frac{16 \sqrt{2}  \mathit{C}_2
    (9 r^4+2 r^2-1) }{3 m_{b}^3 r^2 (r^2-1)^3}-\frac{16 \sqrt{2}  \mathit{C}_3  (r^4+1) }{3
   m_{b}^3 r^2 (r^2-1)^3}+\frac{8 \sqrt{2}   (26 r^4-4 r^2-2) }{3 m_{b}^5 r^2 (r^2-1)^4}],
\end{align}
\begin{align}
f'=\alpha_{s}^2e_lC_A C_F&[\frac{8 \sqrt{2}  \mathit{B}_1  (r^4+2 r^2-1) }{3 m_{b}^5 r^2 (r^2-1)^4}+\frac{8 \sqrt{2}  \mathit{B}_2
    (7 r^4-6 r^2+1) }{3 m_{b}^5 r^2 (r^2-1)^4}-\frac{32 \sqrt{2}  \mathit{B}_3  (2 r^2-1) }{3
   m_{b}^5 (r^2-1)^4}+\frac{32 \sqrt{2}  \mathit{C}_1  }{3 m_{b}^3 r^2 (r^2-1)^3}\nonumber\\&+\frac{16 \sqrt{2}
   \mathit{C}_2  (4 r^4-6 r^2+1) }{3 m_{b}^3 r^2 (r^2-1)^3}+\frac{16 \sqrt{2}  \mathit{C}_3  (2 r^4-1)
   }{3 m_{b}^3 r^2 (r^2-1)^3}-\frac{16 \sqrt{2}   (7 r^4-6 r^2+1) }{3 m_{b}^5 r^2 (r^2-1)^4}],
\end{align}

\begin{align}
g'=\alpha_{s}^2e_lC_A C_F&[\frac{64  \mathit{B}_1  (8 r^8+83 r^6-32 r^4-41 r^2+2) r }{3 m_{b}^2 (r^2-1)^4 (2 r^6+22 r^4+10
   r^2+1)}+\frac{64  \mathit{B}_2  (24 r^{10}+232 r^8-275 r^6-26 r^4+79 r^2-14) r }{3 m_{b}^2 (r^2-1)^4
   (2 r^6+22 r^4+10 r^2+1)}\nonumber\\&-\frac{128  \mathit{B}_3  (25 r^8+280 r^6+102 r^4-21 r^2+13) r }{9 m_{b}^2
   (r^2-1)^3 (2 r^6+22 r^4+10 r^2+1)}-\frac{128  \mathit{B}_4  (11 r^{10}+105 r^8-110 r^6+36 r^4+23 r^2-5) r }{9 m_{b}^2 (r^2-1)^4 (2 r^6+22 r^4+10 r^2+1)}\nonumber\\&+\frac{128  \mathit{C}_1  (4 r^{10}+48 r^8+66 r^6+41 r^4+4
   r^2-3) r }{3 (r^2-1)^2 (2 r^8+24 r^6+32 r^4+11 r^2+1)}+\frac{128  \mathit{C}_2  (4 r^8+44 r^6+39 r^4-20
   r^2+3) r }{3 (r^2-1)^2 (2 r^6+22 r^4+10 r^2+1)}\nonumber\\&+\frac{128  \mathit{C}_3  (4 r^8+44 r^6-37 r^4-4
   r^2+3) r }{3 (r^2-1)^2 (2 r^6+22 r^4+10 r^2+1)}-\frac{512  \mathit{C}_5  (r^{12}+11 r^{10}+5 r^8-3 r^4+5
   r^2+1) r^3 }{3 (r^2-1)^4 (2 r^8+24 r^6+32 r^4+11 r^2+1)}\nonumber\\&-\frac{64   (28 r^{10}+276 r^8-297 r^6-125
   r^4+77 r^2+1) r }{3 m_{b}^2 (r^2-1)^4 (2 r^6+22 r^4+10 r^2+1)}],
\end{align}
\begin{align}
h'=\alpha_{s}^2e_lC_A C_F&[\frac{32  \mathit{B}_1  (2 r^{10}-2 r^8+41 r^6-52 r^4-31 r^2+2) r }{3 m_{b}^4 (r^2-1)^5 (2 r^6+22 r^4+10
   r^2+1)}-\frac{32  \mathit{B}_2  (82 r^{10}-150 r^8+161 r^6-26 r^4-41 r^2+14) r }{3 m_{b}^4 (r^2-1)^5
   (2 r^6+22 r^4+10 r^2+1)}\nonumber\\&+\frac{64  \mathit{B}_3  (131 r^8-7 r^6-6 r^4-13) r }{9 m_{b}^4 (r^2-1)^4
   (2 r^6+22 r^4+10 r^2+1)}-\frac{64  \mathit{B}_4  (11 r^{10}+84 r^8-179 r^6-33 r^4+2 r^2-5) r }{9 m_{b}^4
   (r^2-1)^5 (2 r^6+22 r^4+10 r^2+1)}\nonumber\\&+\frac{64  \mathit{C}_1  (4 r^{10}+42 r^8+44 r^6+17 r^4-4 r^2-3) r }{3
   m_{b}^2 (r^2-1)^3 (2 r^8+24 r^6+32 r^4+11 r^2+1)}+\frac{64  \mathit{C}_2  (28 r^8-116 r^6-69 r^4-21 r^2+3) r
   }{3 m_{b}^2 (r^2-1)^3 (2 r^6+22 r^4+10 r^2+1)}\nonumber\\&-\frac{64  \mathit{C}_3  (32 r^8-24 r^6-9 r^4-3 r^2-1) r
   }{m_{b}^2 (r^2-1)^3 (2 r^6+22 r^4+10 r^2+1)}-\frac{256  \mathit{C}_5  (r^{10}+10 r^8-4 r^6-20 r^4-23
   r^2-4) r^5 }{3 m_{b}^2 (r^2-1)^5 (2 r^8+24 r^6+32 r^4+11 r^2+1)}\nonumber\\&+\frac{32   (142 r^{10}-148 r^8+69
   r^6-63 r^4-79 r^2-1) r }{3 m_{b}^4 (r^2-1)^5 (2 r^6+22 r^4+10 r^2+1)}],
\end{align}
\begin{align}
x'=&\alpha_{s}^2e_lC_A C_F[\frac{32  \mathit{B}_1  (8 r^{10}+115 r^8+179 r^6-113 r^4+9 r^2+2) r }{3 m_{b}^4 (r^2-1)^6 (2 r^6+22 r^4+10
   r^2+1)}+\frac{64
    \mathit{C}_3  (36 r^{10}+404 r^8-249 r^6-97 r^4+5 r^2+1) r }{3 m_{b}^2 (r^2-1)^4 (2 r^6+22 r^4+10
   r^2+1)}\nonumber\\&+\frac{32  \mathit{B}_2  (56 r^{12}+640 r^{10}-27 r^8-797 r^6+333 r^4-3 r^2-2) r }{3 m_{b}^4
   (r^2-1)^6 (2 r^6+22 r^4+10 r^2+1)}-\frac{64  \mathit{B}_3  (73 r^{10}+913 r^8+858 r^6-183 r^4+16 r^2+3) r }{9 m_{b}^4 (r^2-1)^5 (2 r^6+22 r^4+10 r^2+1)}\nonumber\\&-\frac{64  \mathit{B}_4  (11 r^{12}+132 r^{10}+187 r^8+114 r^6+131
   r^4+22 r^2+3) r }{9 m_{b}^4 (r^2-1)^6 (2 r^6+22 r^4+10 r^2+1)}-\frac{64
   \mathit{C}_2  (4 r^{10}-4 r^8-675 r^6-15 r^4-9 r^2-1) r }{3 m_{b}^2 (r^2-1)^4 (2 r^6+22 r^4+10 r^2+1)}\nonumber\\&+\frac{64  \mathit{C}_1  (4 r^{12}+44
   r^{10}+34 r^8+131 r^6+17 r^4-27 r^2-3) r }{3 m_{b}^2 (r^2-1)^4 (2 r^8+24 r^6+32 r^4+11 r^2+1)}\nonumber\\&-\frac{256  \mathit{C}_5  (r^{14}+12 r^{12}+24 r^{10}+89 r^8+13 r^6+46 r^4+14 r^2+1) r^3 }{3 m_{b}^2
   (r^2-1)^6 (2 r^8+24 r^6+32 r^4+11 r^2+1)}\nonumber\\&-\frac{32   (80 r^{12}+904 r^{10}-233 r^8-1200 r^6+78 r^4-24 r^2-5) r
   }{3 m_{b}^4 (r^2-1)^6 (2 r^6+22 r^4+10 r^2+1)}],
\end{align}
\begin{align}
y'=& \alpha_{s}^2e_lC_A C_F[-\frac{32 \mathit{B}_1  (2 r^{12}+12 r^{10}-9 r^8-63 r^6-63 r^4+19 r^2+2) r }{3 m_{b}^4 (r^2-1)^6 (2 r^6+22
   r^4+10 r^2+1)}+\frac{64
    \mathit{C}_3  (30 r^{10}-24 r^8+41 r^6+r^4+r^2+1) r }{3 m_{b}^2 (r^2-1)^4 (2 r^6+22 r^4+10
   r^2+1)}\nonumber\\&+\frac{32  \mathit{B}_2  (50 r^{12}+184 r^{10}-149 r^8+91 r^6-119 r^4+41 r^2+2) r }{3 m_{b}^4
   (r^2-1)^6 (2 r^6+22 r^4+10 r^2+1)}-\frac{64  \mathit{B}_3  (83 r^{10}+452 r^8+213 r^6+132 r^4-37 r^2-3) r }{9 m_{b}^4 (r^2-1)^5 (2 r^6+22 r^4+10 r^2+1)}\nonumber\\&+\frac{64  \mathit{B}_4  (11 r^{12}+111 r^{10}-29 r^8-312 r^6-85
   r^4+r^2+3) r }{9 m_{b}^4 (r^2-1)^6 (2 r^6+22 r^4+10 r^2+1)}-\frac{64  \mathit{C}_1  (4 r^{12}+38
   r^{10}-30 r^8-11 r^6-63 r^4-35 r^2-3) r }{3 m_{b}^2 (r^2-1)^4 (2 r^8+24 r^6+32 r^4+11 r^2+1)}\nonumber\\&+\frac{64  \mathit{C}_2
    (14 r^{10}+280 r^8+45 r^6+41 r^4-27 r^2-3) r }{3 m_{b}^2 (r^2-1)^4 (2 r^6+22 r^4+10 r^2+1)}+\frac{256  \mathit{C}_5  (r^{12}+11 r^{10}+8 r^8+12 r^6-99 r^4-31 r^2-2) r^5 }{3 m_{b}^2 (r^2-1)^6
   (2 r^8+24 r^6+32 r^4+11 r^2+1)}\nonumber\\&-\frac{32   (90 r^{12}+414 r^{10}-119 r^8-332 r^6-280 r^4+22 r^2+5) r }{3
   m_{b}^4 (r^2-1)^6 (2 r^6+22 r^4+10 r^2+1)}],
\end{align}
where we have the one-loop scalar Passarino-Veltman integrals
\begin{align}
&\mathit{B}_1=\mathit{B}_0(0,m_{b}^2,m_{b}^2),\\
&\mathit{B}_2=\mathit{B}_0(m_{b}^2 (-1+2 r^2),0,m_{b}^2),\\
&\mathit{B}_3=\mathit{B}_0(0,m_{b}^2 r^2,m_{b}^2 r^2),\\
&\mathit{B}_4=\mathit{B}_0(4 m_{b}^2 r^2,0,0),\\
&\mathit{C}_1=\mathit{C}_0(m_{b}^2,0,m_{b}^2 (-1+2 r^2),0,m_{b}^2,m_{b}^2),\\
&\mathit{C}_2=\mathit{C}_0(m_{b}^2 r^2,m_{b}^2 (-1+2 r^2),0,m_{b}^2,0,m_{b}^2 r^2),\\
&\mathit{C}_3=\mathit{C}_0(m_{b}^2 r^2,m_{b}^2 (-1+2 r^2),0,0,m_{b}^2 r^2,m_{b}^2),\\
&\mathit{C}_4=\mathit{C}_0(4 m_{b}^2 r^2,m_{b}^2,m_{b}^2 (-1+2 r^2),m_{b}^2,0,0),\\
&\mathit{C}_5=\mathit{C}_0(m_{b}^2,4 m_{b}^2 r^2,m_{b}^2 (-1+2 r^2),0,0,m_{b}^2),\\
&\mathit{C}_6=\mathit{C}_0(m_{b}^2,4 m_{b}^2 r^2,m_{b}^2 (-1+2 r^2),m_{b}^2,0,0)
\end{align}

The fragmentation functions for heavy quarkonium can be found in Refs.~\cite{Braaten:1993mp,Yuan:1994hn}. Here we list some
 charmonium fragmentation probabilities
\begin{align}
&\int_{0}^{1} dz\,\mathit{D}_{c \to \eta_{c}}(z,3m_{c})=\frac{8}{27\pi}\alpha_{s}(2m_{c})^2  \frac{\left|\mathit{R}(0) \right|^2}{m_{c}^3} \left[\frac{773}{30}-37\ln2\right],\\
&\int_{0}^{1} dz\,\mathit{D}_{c \to \chi_{c0}}(z,3m_{c})=\frac{4}{81\pi}\alpha_{s}(2m_{c})^2 \frac{\left|\mathit{R}_{P}^{'}(0)\right|^2}{m_{c}^5} \left[\frac{119617}{35}-4926\ln2 \right],\\
&\int_{0}^{1} dz\,\mathit{D}_{c \to \chi_{c1}}(z,3m_{c})=\frac{32}{27\pi}\alpha_{s}(2m_{c})^2 \frac{\left|\mathit{R}_{P}^{'}(0)\right|^2}{m_{c}^5} \left[\frac{1151}{7}-237\ln2 \right],\\
&\int_{0}^{1} dz\,\mathit{D}_{c \to \chi_{c2}}(z,3m_{c})=\frac{16}{81\pi}\alpha_{s}(2m_{c})^2 \frac{\left|\mathit{R}_{P}^{'}(0)\right|^2}{m_{c}^5} \left[\frac{54743}{35}-2256\ln2 \right].
\end{align}

 \end{widetext}


\begin{thebibliography}{99}
\bibitem{Katrenko:2019vdd}
  P.~Katrenko {\it et al.} [Belle Collaboration],
  Phys.\ Rev.\ Lett.\  {\bf 124}, no. 12, 122001 (2020)
  doi:10.1103/PhysRevLett.124.122001
  [arXiv:1910.10915 [hep-ex]].


\bibitem{Shen:2010iu}
  C.~P.~Shen {\it et al.} [Belle Collaboration],
  Phys.\ Rev.\ D {\bf 82}, 051504 (2010)
  doi:10.1103/PhysRevD.82.051504
  [arXiv:1008.1774 [hep-ex]].


\bibitem{Brambilla:2010cs}
  N.~Brambilla {\it et al.},
  Eur.\ Phys.\ J.\ C {\bf 71}, 1534 (2011)
  doi:10.1140/epjc/s10052-010-1534-9
  [arXiv:1010.5827 [hep-ph]].


\bibitem{Fulton:1988ug}
  R.~Fulton {\it et al.} [CLEO Collaboration],
  Phys.\ Lett.\ B {\bf 224}, 445 (1989).
  doi:10.1016/0370-2693(89)91476-7


\bibitem{Briere:2004ug}
  R.~A.~Briere {\it et al.} [CLEO Collaboration],
  Phys.\ Rev.\ D {\bf 70}, 072001 (2004)
  doi:10.1103/PhysRevD.70.072001
  [hep-ex/0407030].


\bibitem{He:2009by}
  Z.~G.~He and J.~X.~Wang,
  Phys.\ Rev.\ D {\bf 81}, 054030 (2010)
  doi:10.1103/PhysRevD.81.054030
  [arXiv:0911.0139 [hep-ph]].


\bibitem{He:2019rwt}
  Z.~G.~He, B.~A.~Kniehl and X.~P.~Wang,
  Phys.\ Rev.\ D {\bf 101}, no. 7, 074002 (2020)
  doi:10.1103/PhysRevD.101.074002
  [arXiv:1912.10232 [hep-ph]].


\bibitem{Jia:2020csg}
  S.~Jia, X.~Zhou and C.~Shen,
  Front.\ Phys.\ (Beijing) {\bf 15}, no. 6, 64301 (2020)
  doi:10.1007/s11467-020-0978-0
  [arXiv:2005.05892 [hep-ex]].


\bibitem{Bodwin:1994jh}
  G.~T.~Bodwin, E.~Braaten and G.~P.~Lepage,
  Phys.\ Rev.\ D {\bf 51}, 1125 (1995)
  Erratum: [Phys.\ Rev.\ D {\bf 55}, 5853 (1997)]
  doi:10.1103/PhysRevD.55.5853, 10.1103/PhysRevD.51.1125
  [hep-ph/9407339].


\bibitem{Jia:2007hy}
  Y.~Jia,
  Phys.\ Rev.\ D {\bf 76}, 074007 (2007)
  doi:10.1103/PhysRevD.76.074007
  [arXiv:0706.3685 [hep-ph]].


\bibitem{Gong:2008ue}
  B.~Gong, Y.~Jia and J.~X.~Wang,
  Phys.\ Lett.\ B {\bf 670}, 350 (2009)
  doi:10.1016/j.physletb.2008.10.063
  [arXiv:0808.1034 [hep-ph]].


\bibitem{Braguta:2009xu}
  V.~V.~Braguta and V.~G.~Kartvelishvili,
  Phys.\ Rev.\ D {\bf 81}, 014012 (2010)
  doi:10.1103/PhysRevD.81.014012
  [arXiv:0907.2772 [hep-ph]].


\bibitem{Braguta:2010zz}
  V.~V.~Braguta, A.~K.~Likhoded and A.~V.~Luchinsky,
  Phys.\ Atom.\ Nucl.\  {\bf 73}, 1054 (2010)
  [Yad.\ Fiz.\  {\bf 73}, 1091 (2010)].
  doi:10.1134/S1063778810060207


\bibitem{Kang:2007uv}
  D.~Kang, T.~Kim, J.~Lee and C.~Yu,
  Phys.\ Rev.\ D {\bf 76}, 114018 (2007)
  doi:10.1103/PhysRevD.76.114018
  [arXiv:0707.4056 [hep-ph]].


\bibitem{Zhang:2008pr}
  Y.~J.~Zhang and K.~T.~Chao,
  Phys.\ Rev.\ D {\bf 78}, 094017 (2008)
  doi:10.1103/PhysRevD.78.094017
  [arXiv:0808.2985 [hep-ph]].


\bibitem{Sang:2012yh}
  W.~L.~Sang, H.~T.~Chen and Y.~Q.~Chen,
  Phys.\ Rev.\ D {\bf 86}, 114004 (2012)
  doi:10.1103/PhysRevD.86.114004
  [arXiv:1209.5270 [hep-ph]].


\bibitem{Chen:2012zzg}
  H.~T.~Chen, W.~L.~Sang and P.~Wu,
  Commun.\ Theor.\ Phys.\  {\bf 57}, 665 (2012).
  doi:10.1088/0253-6102/57/4/22


\bibitem{Li:2020ggh}
  S.~Y.~Li, Z.~Y.~Li, Z.~G.~Si, Z.~J.~Yang and X.~Zhang,
  arXiv:2007.07706 [hep-ph].


\bibitem{Zhu:2015jha}
  R.~Zhu,
  Phys.\ Rev.\ D {\bf 92}, no. 7, 074017 (2015)
  doi:10.1103/PhysRevD.92.074017
  [arXiv:1507.02031 [hep-ph]].


\bibitem{Zhu:2015qoa}
  R.~Zhu,
  JHEP {\bf 1509}, 166 (2015)
  doi:10.1007/JHEP09(2015)166
  [arXiv:1508.01445 [hep-ph]].


\bibitem{Hao:2006nf}
  G.~Hao, Y.~Jia, C.~F.~Qiao and P.~Sun,
  JHEP {\bf 0702}, 057 (2007)
  doi:10.1088/1126-6708/2007/02/057
  [hep-ph/0612173].


\bibitem{Gao:2007fv}
  Y.~J.~Gao, Y.~J.~Zhang and K.~T.~Chao,
  hep-ph/0701009.


\bibitem{Zhu:2017lwi}
  R.~Zhu,
  Nucl.\ Phys.\ B {\bf 931}, 359 (2018)
  doi:10.1016/j.nuclphysb.2018.04.018
  [arXiv:1710.07011 [hep-ph]].


\bibitem{Zhu:2017lqu}
  R.~Zhu, Y.~Ma, X.~L.~Han and Z.~J.~Xiao,
  Phys.\ Rev.\ D {\bf 95}, no. 9, 094012 (2017)
  doi:10.1103/PhysRevD.95.094012
  [arXiv:1703.03875 [hep-ph]].


\bibitem{Petrelli:1997ge}
  A.~Petrelli, M.~Cacciari, M.~Greco, F.~Maltoni and M.~L.~Mangano,
  Nucl.\ Phys.\ B {\bf 514}, 245 (1998)
  doi:10.1016/S0550-3213(97)00801-8
  [hep-ph/9707223].


\bibitem{Chao:2012iv}
  K.~T.~Chao, Y.~Q.~Ma, H.~S.~Shao, K.~Wang and Y.~J.~Zhang,
  Phys.\ Rev.\ Lett.\  {\bf 108}, 242004 (2012)
  doi:10.1103/PhysRevLett.108.242004
  [arXiv:1201.2675 [hep-ph]].


\bibitem{Braaten:1993mp}
  E.~Braaten, K.~m.~Cheung and T.~C.~Yuan,
  Phys.\ Rev.\ D {\bf 48}, 4230 (1993)
  doi:10.1103/PhysRevD.48.4230
  [hep-ph/9302307].


\bibitem{Yuan:1994hn}
  T.~C.~Yuan,
  Phys.\ Rev.\ D {\bf 50}, 5664 (1994)
  doi:10.1103/PhysRevD.50.5664
  [hep-ph/9405348].


\bibitem{Qiao:2011yk}
  C.~F.~Qiao, L.~P.~Sun, D.~S.~Yang and R.~L.~Zhu,
  Eur.\ Phys.\ J.\ C {\bf 71}, 1766 (2011)
  doi:10.1140/epjc/s10052-011-1766-3
  [arXiv:1103.1106 [hep-ph]].


\bibitem{Tanabashi:2018oca}
  M.~Tanabashi {\it et al.} [Particle Data Group],
  Phys.\ Rev.\ D {\bf 98}, no. 3, 030001 (2018).
  doi:10.1103/PhysRevD.98.030001


\bibitem{Qiao:2012vt}
  C.~F.~Qiao and R.~L.~Zhu,
  Phys.\ Rev.\ D {\bf 87}, no. 1, 014009 (2013)
  doi:10.1103/PhysRevD.87.014009
  [arXiv:1208.5916 [hep-ph]].


\bibitem{Qiao:2012hp}
  C.~F.~Qiao, P.~Sun, D.~Yang and R.~L.~Zhu,
  Phys.\ Rev.\ D {\bf 89}, no. 3, 034008 (2014)
  doi:10.1103/PhysRevD.89.034008
  [arXiv:1209.5859 [hep-ph]].


\bibitem{Qiao:2014pfa}
  C.~F.~Qiao and R.~L.~Zhu,
  Phys.\ Rev.\ D {\bf 89}, no. 7, 074006 (2014)
  doi:10.1103/PhysRevD.89.074006
  [arXiv:1403.1918 [hep-ph]].


\bibitem{Zhu:2016udl}
  R.~Zhu and J.~P.~Dai,
  Phys.\ Rev.\ D {\bf 94}, no. 9, 094034 (2016)
  doi:10.1103/PhysRevD.94.094034
  [arXiv:1610.00288 [hep-ph]].


\bibitem{Patel:2015tea}
  H.~H.~Patel,
  Comput.\ Phys.\ Commun.\  {\bf 197}, 276 (2015)
  doi:10.1016/j.cpc.2015.08.017
  [arXiv:1503.01469 [hep-ph]].


\bibitem{Bodwin:2005gg}
  G.~T.~Bodwin, J.~Lee and D.~K.~Sinclair,
  Phys.\ Rev.\ D {\bf 72}, 014009 (2005)
  doi:10.1103/PhysRevD.72.014009
  [hep-lat/0503032].

\bibitem{Schuler:1994hy}
G.~A.~Schuler,
[arXiv:hep-ph/9403387 [hep-ph]].

\end{thebibliography}
\end{document}